\documentclass[aps,onecolumn,showpacs,nofootinbib,showkeys]{revtex4-2}
\usepackage[margin=3.15cm]{geometry}

\RequirePackage[T1]{fontenc}

\usepackage{epsfig,graphicx,amsmath,amssymb,bm}
\RequirePackage{mathptmx}      
\usepackage{mathrsfs}
\usepackage{amssymb}
\RequirePackage{color}
\usepackage{changes}
\usepackage{slashed}
\usepackage{multirow}
\usepackage{scalerel}
\usepackage{tikz-feynman}
\usepackage{textcomp}
\usepackage{subcaption}
\usepackage[english]{babel}
\usepackage{float}
\RequirePackage{hyperref}
\hypersetup{
    linktocpage,
    colorlinks,
    citecolor=blue,
    filecolor=black,
    linkcolor=blue,
    urlcolor=blue,
}
\usepackage{nccmath}

\begin{document}

\title{Decays $f_1 \to a_0 \pi,\pi \pi \pi(\eta)$ and $f_1 \to K K \pi$ in the chiral $U(3) \times U(3)$ quark NJL model }


\author{M. K. Volkov$^{1}$}\email{volkov@theor.jinr.ru}
\author{A. A. Pivovarov$^{1}$}\email{pivovarov@theor.jinr.ru}
\author{K. Nurlan$^{1,2,3}$}\email{nurlan@theor.jinr.ru}

\affiliation{$^1$ Bogoliubov Laboratory of Theoretical Physics, JINR, 
                 141980 Dubna, Moscow region, Russia \\
                $^2$ The Institute of Nuclear Physics, Almaty, 050032, Kazakhstan}   


\begin{abstract}
The branching fractions of the axial vector meson decays $f_1 \to a_0 \pi,\pi\pi\eta$, $f_1 \to \rho \pi,\pi\pi\pi$ and $f_1 \to KK\pi$ are calculated in the standard $U(3) \times U(3)$ quark NJL model. The intermediate channels with the states $a_0(980)\pi$ and $f_0(500)\eta$ are taken into account in the decay $f_1 \to \pi\pi\eta$. In the case of the scalar mesons, the $\bar{q}q$ representation as a chiral symmetric partners of the pseudoscalar mesons is used. It is shown that the decays $f_1 \to \rho \pi$ and $f_1 \to3\pi$ occur due to the mass difference of the $u$ and $d$ quarks. All the results are obtained without using any additional arbitrary parameters and are in satisfactory agreement with the known experimental data.


\end{abstract}

\pacs{}

\maketitle
\section{Introduction}
    Axial vector mesons play an important role in the processes of hadron interactions at low energies. The theoretical analysis of various mesonic $\tau$ lepton decays with strange and non-strange mesons in the final state showed that axial vector channels are dominant in the calculation of decay widths ~\cite{Volkov:2017arr,Guo:2008sh,Dai:2020vfc,K:2023kgj,Volkov:2023pmy}. In particular, in the $\tau$ lepton decay into three $\pi$ mesons, the main contribution is given by the axial vector channel with the intermediate meson $a_1(1260)$ \cite{Colangelo:1996hs,GomezDumm:2003ku,JPAC:2018zwp,Dumm:2009va,Nugent:2013hxa,Nugent:2013hxa,Sadasivan:2021emk,Volkov:2024wsu}. At the same time, the $\tau$ decays with the production of the axial vector meson $\tau \to f_1(1285) \pi \nu_\tau$ are intensively investigated from both theoretical and experimental points of view \cite{ParticleDataGroup:2024cfk,Calderon:2012zi,Volkov:2018fyv,Oset:2018zgc}. The properties and decays of the axial vector meson $f_1(1285)$ were studied based on the $\pi$ meson scattering on nucleons $\pi^- N \to N \pi^- (\pi^+\pi^-\pi^0,\pi^+\pi^- \eta)$~\cite{Dorofeev:2011zz}. Based on the analysis of experimental data, the estimations for the decay widths ratio $\Gamma (f_1 \to \pi^+\pi^-\pi^0) / \Gamma (f_1 \to \pi^+\pi^-\eta)$ and the restrictions for the branching fraction $Br(f_1 \to \pi\rho) < 0.31 \%$ were obtained. The recent investigations of the $f_1$ meson production in the $pp$ collisions have been carried out by the  LHCb and ALICE collaborations at CERN~\cite{LHCb:2013ged,ALICE:2024rjz}. 

    For theoretical study of low energy processes of meson interactions various phenomenological models are applied, which are usually based on the approximate chiral symmetry of strong interactions. One of the successful models of this type is the Nambu--Jona-Lasinio (NJL) model. The NJL model \cite{Nambu:1961tp,Eguchi:1976iz,Ebert:1982pk,Volkov:1984kq,Volkov:1986zb,Ebert:1985kz,Vogl:1991qt,Klevansky:1992qe,Volkov:1993jw,Hatsuda:1994pi,Ebert:1994mf,Buballa:2003qv,Volkov:2005kw} is a phenomenological quark model based on the chiral group $U(3)\times U(3)$ that has successfully proved itself in describing the processes of meson interactions at low energies. 

    Significant progress has been achieved in the study of the anomalous decays of the axial vector mesons $a_1(1260)$ and $f_1(1285)$~\cite{Amelin:1994ii,Lutz:2008km,Osipov:2017ray,Osipov:2018ejk,Osipov:2018iah,Xie:2019iwz}. In particular, within the NJL model the effective Lagrangians for the vertices AVV, AAAV and the widths of the anomalous decays $f_1 \to \rho(\omega)\gamma$, $f_1 \to \pi \pi \gamma$, $f_1 \to \rho\pi\pi$ and $a_1 \to \omega \pi \pi$ were obtained in agreement with the known experimental data ~\cite{Osipov:2017ray,Osipov:2018ejk,Osipov:2018iah}. The theoretical prediction for the decay $f_1 \to \rho\gamma$ is compatible the recent data from the CLAS collaboration~\cite{CLAS:2016zjy} which deviate significantly from the average PDG data~\cite{ParticleDataGroup:2024cfk}.

    In the present work, the decays of the axial vector meson $f_1(1285)$ are considered in the framework of the NJL model. The branching fraction of the decay $f_1 \to a_0 \pi$ is calculated, and the dominant role of the contribution of the scalar mesons in the description the process $f_1 \to \pi \pi \eta$ is demonstrated. The decays $f_1 \to \rho\pi$ and $f_1 \to 3\pi$ that occur due to isospin breaking are also considered. It is shown that the branching fraction is mostly determined by the triangle quark diagram $f_1 \to \rho^{\pm} \pi^{\mp}$ occurring due to the difference of the mass of the $u$ and $d$ quarks. Besides, the decay of the $f_1$ meson with the strange particles $f_1 \to KK\pi$ is considered. 

    The paper is organized in the following way. The Lagrangian of the quark-meson interactions in the NJL model is presented in Section~\ref{section::2}. In Section~\ref{section::3} the decays $f_1 \to a_0 \pi$ and $f_1 \to \pi \pi \eta$ are considered. In Section~\ref{section::4} the suppressed decays $f_1 \to \rho \pi$ and $f_1 \to \pi \pi \pi$ are described. In Section~\ref{section5} the amplitude and branching fractions of the decays with the strange mesons $f_1 \to KK\pi$ are given. The discussion of the results is given in Conclusion.

\section{Lagrangian of the NJL model}
\label{section::2}
The NJL model is formulated based on the QCD motivated local 4-quark chiral interaction in the minimal order in $1/N_c$ (the mean field approximation). The Lagrangian has been obtained after the bosonization and renormalization in the one-loop quark approximation. The fragment of the NJL model Lagrangian containing the quark-meson vertices needed for our calculations has the following form~\cite{Volkov:1986zb,Volkov:1993jw,Volkov:2005kw}:
\begin{eqnarray}
	\Delta L_{int} & = & \bar{q}\left\{\sum_{i=0,\pm}\left[ig_{\pi}\gamma^{5}\lambda^{\pi}_{i}\pi^{i} +
	ig_{K}\gamma^{5}\lambda^{K}_{i}K^{i} + \frac{g_{\rho}}{2}\gamma^{\mu}\lambda^{\rho}_{i}\rho^{i}_{\mu} + \frac{g_{a_1}}{2}\gamma^{\mu}\gamma^{5}\lambda^{a_1}_{i}a^{i}_{1\mu} + \frac{g_{K^{*}}}{2}\gamma^{\mu}\lambda^{K}_{i}K^{*i}_{\mu} + g_{\sigma} \lambda^{a_0}_i a_0^i + g_{K_0^*}\lambda^{K}_{i}K_0^{*i}\right] \right. \nonumber\\
	&& + ig_{K}\gamma^{5}\lambda_{0}^{\bar{K}}\bar{K}^{0} + \frac{g_{K^{*}}}{2}\gamma^{\mu}\lambda^{\bar{K}}_{0}\bar{K}^{*0}_{\mu} + g_{K_0^*}\lambda_{0}^{\bar{K}}\bar{K}_0^{*0} + \frac{1}{2} \gamma^\mu \gamma^5 \left(g_{f_1}^u\lambda_u^{f_1}\cos{\phi} - g_{f_1}^s\lambda_s^{f_1}\sin{\phi}\right) f_{1\mu} \nonumber\\
    && + \frac{1}{2} \gamma^\mu \gamma^5 \left(g_{f_1}^u\lambda_u^{f_1}\sin{\phi} + g_{f_1}^s\lambda_s^{f_1}\cos{\phi}\right) f_{1\mu}' + \left(g_\sigma \lambda_u^{f_0}\cos{\bar{\theta}_{\sigma}} - g_{\sigma}^s\lambda_s^{f_0}\sin{\bar{\theta}_{\sigma}}\right) f_0 \nonumber\\
    && \left. + i \gamma^5 \left(g_{\eta}^u\lambda_u^{\eta}\sin{\bar{\theta}_{\eta}} + g_{\eta}^s\lambda_s^{\eta}\cos{\bar{\theta}_{\eta}}\right) \eta \right\}q,
\end{eqnarray}
where $q = \left(u, d, s\right)$ is a quark triplet, $\lambda$ are the linear combinations of the Gell-Mann matrices, the isoscalar axial vector mesons $f_1 = f_1(1285)$ and $f_1' = f_1(1420)$ are considered as mixed states, $\phi = 24^\circ$ is the mixing angle of the $f_1$ mesons~\cite{LHCb:2013ged}, $\bar{\theta}_{\sigma} = \theta_0 - \theta_\sigma$ is the mixing angle of the $f_0$ mesons~\cite{Volkov:1998ax} and $\bar{\theta}_{\eta} = \theta_0 - \theta_\eta$ is the mixing angle of the $\eta$ mesons~\cite{Volkov:1998ax}, where $\theta_0 = 35.3^\circ$ is the ideal mixing angle and $\theta_\sigma = 24^\circ$, $\theta_\eta = -19^\circ$.

The scalar mesons are considered here as quark-antiquark states, being the chiral partners of the pseudoscalar mesons.

The coupling constants of the mesons with quarks are the result of the renormalisation of the Lagrangian in the one-loop quark approximation:
\begin{eqnarray}
	&g_{\sigma} = \sqrt{\frac{1}{4 I_{20}}}, \quad g_{\eta}^u \approx g_{\pi} = \sqrt{\frac{Z_{\pi}}{4 I_{20}}}, \quad g_{\rho} = g_{a_1} = g_{f_1}^u = \sqrt{\frac{3}{2 I_{20}}}, \quad g_{K} = \sqrt{\frac{Z_{K}}{4 I_{11}}},& \nonumber\\    
    &g_{\sigma}^s = \sqrt{\frac{1}{4 I_{02}}}, \quad g_{\eta}^s = \sqrt{\frac{Z_{s}}{4 I_{02}}}, \quad  g_{f_1}^s = \sqrt{\frac{3}{2 I_{02}}}, \quad g_{K^{*}} = \sqrt{\frac{3}{2 I_{11}}}, \quad g_{K_0^*} = \sqrt{\frac{1}{4 I_{11}}}&
\end{eqnarray}
where $Z_\pi$, $Z_s$ and $Z_K$ are the additional renormalization constants appearing as a result of taking into account transitions between the pseudoscalar and axial vector states:
\begin{eqnarray}
	&Z_{\pi} = \left(1 - 6\frac{m^{2}_{u}}{M^{2}_{a_{1}(1260)}}\right)^{-1}, \quad
	Z_{K} = \left(1 - \frac{3}{2}\frac{(m_{u} + m_{s})^{2}}{M^{2}_{K_{1A}}}\right)^{-1}, \quad Z_{s} = \left(1 - 6\frac{m^{2}_{s}}{M^{2}_{f_{1}(1420)}}\right)^{-1}& \nonumber\\
	&M^{2}_{K_{1A}} = \left(\frac{\sin^{2}{\alpha}}{M^{2}_{K_{1}(1270)}} + \frac{\cos^{2}{\alpha}}{M^{2}_{K_{1}(1400)}}\right)^{-1}.&
\end{eqnarray}
Here $m_u = 270$~MeV and $m_s = 420$~MeV are the constituent masses of the $u$ and $s$ quarks. In most cases, the difference between the masses of the $u$ and $d$ quarks may be neglected. In the cases where it is impossible, the value $m_d - m_u = 4$~MeV can be used. This difference was obtained by analyzing the process $\omega \to \pi^+ \pi^-$~\cite{Volkov:1986zb}. The mixing angle of the meson states $K_{1}(1270)$ and $K_{1}(1400)$ is $\alpha = 57^\circ$.

The integrals $I_{nm}$ appear in quark loops during the renormalization of the Lagrangian and take the following form:
\begin{equation}
\label{integral}
	I_{nm} = -i\frac{N_{c}}{(2\pi)^{4}}\int\frac{\theta(\Lambda^{2} + k^2)}{(m_{u}^{2} - k^2)^{n}(m_{s}^{2} - k^2)^{m}}
	\mathrm{d}^{4}k,
\end{equation}
where $\Lambda = 1265$~MeV is the cut-off parameter~\cite{Volkov:2005kw}.

\section{Decays $f_1 \to a_0 \pi$ and $f_1 \to \pi \pi \eta$}
\label{section::3}
\begin{figure*}[t]
 \centering
  \begin{subfigure}{0.5\textwidth}
   \centering
   \begin{tikzpicture}
    \begin{feynman} 
      \vertex (a) {\(f_1 \)};
      \vertex [dot, right=1.2cm of a] (b) {};
      \vertex [dot, above right=1.6cm of b] (c) {};
      \vertex [dot, below right=1.6cm of b] (d) {};
      \vertex [right=1.2cm of c] (g) {\(\pi\)};
      \vertex [right=1.2cm of d] (f) {\(a_0\)};
      \diagram* {
        (a) -- [double] (b),
        (b) -- [fermion] (c),
        (c) -- [fermion] (d),
        (d) -- [fermion] (b),  
        (c) -- [double] (g),         
        (d) -- [double] (f),
      };
     \end{feynman}
    \end{tikzpicture}
  \end{subfigure}%
 \centering
 \begin{subfigure}{0.5\textwidth}
  \centering
   \begin{tikzpicture}
     \begin{feynman}
      \vertex (a) {\(f_1 \)};
      \vertex [dot, right=1.2cm of a] (b) {};
      \vertex [dot, above right=1.6cm of b] (c) {};
      \vertex [dot, below right=1.6cm of b] (d) {};
      \vertex [dot, right=1.0cm of c] (g) {};
      \vertex [dot, right=0.8cm of g] (h) {};
      \vertex [right=1.0cm of h] (k) {\(\pi \)};
      \vertex [right=1.2cm of d] (f) {\(a_0\)};
      \diagram* {
        (a) -- [double] (b),
        (b) -- [fermion] (c),
        (c) -- [fermion] (d),
        (d) -- [fermion] (b),
        (c) -- [double, edge label'=\({a_1} \)] (g),
        (g) -- [fermion, inner sep=1pt, half left] (h),
        (h) -- [fermion, inner sep=1pt, half left] (g),         
        (h) -- [double] (k),         
        (d) -- [double] (f),
      };
     \end{feynman}
    \end{tikzpicture}
  \end{subfigure}%
 \caption{Quark diagrams of the decay $f_1 \to a_0 \pi$.}
 \label{diagram1}
\end{figure*}
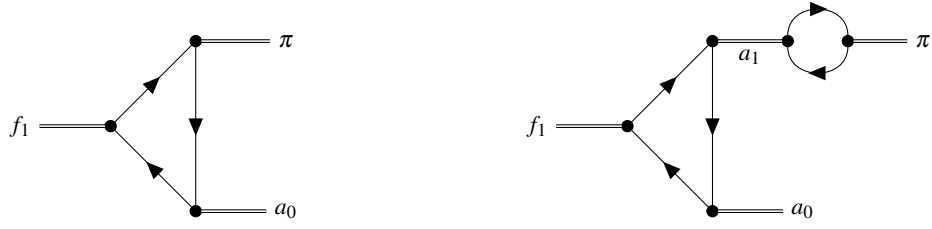

The decay $f_1 \to a_0 \pi$ is described with the quark diagrams presented in Fig.~\ref{diagram1}. The amplitude takes the form

\begin{eqnarray}
\label{a0pi}
    \mathcal{M}(f_1 \to a_0 \pi) & = & -ig_{f_1}^u \sqrt{Z_\pi} \cos{\phi} {\left( p_{a_0}  - A_\pi p_\pi\right)}^\mu e_\mu^*(p_{f_1}),
\end{eqnarray}
where $p_{a_0}$, $p_\pi$ are the momenta of the mesons, $e_\mu(p_{f_1})$ is the polarization vector of the meson $f_1$. The factor $A_\pi = 1- 12m^2_u/M^2_{a_1}$ takes into account the non diagonal $a_1 - \pi$ transition in the final state. The width of the decay $f_1 \to a_0 \pi$ can be calculated by using the formula 
\begin{eqnarray}
\label{a0piwidth}
\Gamma(f_1 \to a_0 \pi) =
\frac{1}{2J_{f_1}+1} \frac{\sqrt{E^2_\pi - M^2_\pi}}{8\pi M^2_{f_1}} \cdot \, {\mid \mathcal{M}(f_1 \to a_0 \pi) \mid}^2,
\end{eqnarray}
where $J_{f_1}=1$, $E_\pi = (M^2_{f_1}+M^2_{\pi}-M^2_{a_0})/2M^2_{f_1}$ is the energy of the $\pi$ meson in the rest system of the $f_1$ meson. 

As a result, the following estimation can be obtained for the branching fraction of the decay $f_1 \to a_0 \pi$
\begin{eqnarray}
Br(f_1 \to a_0 \pi) = (39.4 \pm 5.91) \, \%.
\end{eqnarray}

Calculation uncertainties in the framework of the NJL model can be estimated at a level of $15\%$ \cite{Volkov:2017arr,K:2023kgj,Volkov:2023pmy}. This value is based on a comparison of numerous calculations carried out within the NJL model earlier with the known experimental results.

The obtained estimation of the branching fraction is in agreement with the experimental data given in the PDG~\cite{ParticleDataGroup:2024cfk}
\begin{eqnarray}
Br(f_1 \to a_0 \pi)_{PDG} = 38.0 \pm 4.0 \, \%.
\end{eqnarray}

Similar calculations can be carried out for the decay $f_1 \to f_0 \eta$ where $f_0 = f_0(500)$. In this decay, the transition $\eta-f_1$ takes place instead of the $\pi-a_1$ transition in the amplitude (\ref{a0pi}). The decay is described by the following amplitude:
\begin{eqnarray}
\label{f0eta}
    \mathcal{M}(f_1 \to f_0 \eta) & = &  \left( g^u_{f_1} \sqrt{Z_\pi} \cos{\phi} \cos{\bar{\theta_\sigma}} \sin{\bar{\theta_\eta}} - \sqrt{2} g^s_{f_1} \sqrt{Z_s} \sin{\phi} \sin{\bar{\theta_\sigma}} \cos{\bar{\theta_\eta}} \right)
    {\left( p_{f_0}  - A_\eta p_\eta\right)}^\mu e_\mu^*(p_{f_1}),
\end{eqnarray}
where the constant describing the $\eta-f_1$ transition takes the form
\begin{eqnarray}
\nonumber
A_\eta & = & 1 - 3\sqrt{6}\frac{\left( m_u g_\sigma {\cos{\phi}}^2 \cos{\bar{\theta_\sigma}} + \sqrt{2} m_s g^s_{\sigma} {\sin{\phi}}^2 \sin{\bar{\theta_\sigma}} \right) 
\left( m_u \sqrt{Z_\pi}\cos{\phi} \sin{\bar{\theta_\eta}} - m_s \sqrt{Z_s}\sin{\phi} \cos{\bar{\theta_\eta}}\right)}
{M^2_{f_1(1285)} \cdot \left( g^u_{f_1} \sqrt{Z_\pi} \cos{\phi} \cos{\bar{\theta_\sigma}} \sin{\bar{\theta_\eta}} - \sqrt{2} g^s_{f_1} \sqrt{Z_s} \sin{\phi} \sin{\bar{\theta_\sigma}} \cos{\bar{\theta_\eta}} \right)} \nonumber\\
&& - 3\sqrt{6}\frac{\left( m_u g_\sigma {\cos{\phi}} \sin{\phi} \cos{\bar{\theta_\sigma}} - \sqrt{2} m_s g^s_{\sigma} {\sin{\phi}} \cos{\phi} \sin{\bar{\theta_\sigma}} \right) 
\left( m_u \sqrt{Z_\pi}\sin{\phi} \sin{\bar{\theta_\eta}} + m_s \sqrt{Z_s}\cos{\phi} \cos{\bar{\theta_\eta}}\right)}
{M^2_{f_1(1420)} \cdot \left( g^u_{f_1} \sqrt{Z_\pi} \cos{\phi} \cos{\bar{\theta_\sigma}} \sin{\bar{\theta_\eta}} - \sqrt{2} g^s_{f_1} \sqrt{Z_s} \sin{\phi} \sin{\bar{\theta_\sigma}} \cos{\bar{\theta_\eta}} \right)}.
\end{eqnarray}

As a result, the following theoretical estimation of the branching fraction of the decay $f_1 \to f_0 \eta$ can be obtained in the quark NJL model.
\begin{eqnarray}
Br(f_1 \to f_0 \eta) = (4.2 \pm 0.63) \, \%.
\end{eqnarray}

\begin{figure*}[t]
 \centering
  \begin{subfigure}{0.5\textwidth}
   \centering
   \begin{tikzpicture}
    \begin{feynman} 
      \vertex (a) {\(f_1 \)};
      \vertex [dot, right=1.2cm of a] (b) {};
      \vertex [dot, above right=1.4cm of b] (c) {};
      \vertex [dot, below right=1.4cm of b] (d) {};
      \vertex [right=1.2cm of c] (e) {\(\pi^\pm\)};
      \vertex [dot, right=1.2cm of d] (f) {};
      \vertex [dot, above right=1.4cm of f] (g) {};
      \vertex [dot, below right=1.4cm of f] (h) {};
      \vertex [right=1.2cm of g] (i) {\(\pi^\mp\)};
      \vertex [right=1.2cm of h] (j) {\(\eta\)};
      \diagram* {
        (a) -- [double] (b),
        (b) -- [fermion] (c),
        (c) -- [fermion] (d),
        (d) -- [fermion] (b),  
        (c) -- [double] (e),         
        (d) -- [double, edge label'=\(a^\mp_0\)] (f),
        (f) -- [fermion] (g),
        (h) -- [fermion] (f),
        (g) -- [fermion] (h),
        (g) -- [double] (i),
        (h) -- [double] (j),
      };
     \end{feynman}
    \end{tikzpicture}
  \end{subfigure}%
 \centering
 \begin{subfigure}{0.5\textwidth}
  \centering
   \begin{tikzpicture}
     \begin{feynman}
      \vertex (a) {\(f_1 \)};
      \vertex [dot, right=1.2cm of a] (b) {};
      \vertex [dot, above right=1.4cm of b] (c) {};
      \vertex [dot, below right=1.4cm of b] (d) {};
      \vertex [right=1.2cm of c] (e) {\(\eta\)};
      \vertex [dot, right=1.2cm of d] (f) {};
      \vertex [dot, above right=1.4cm of f] (g) {};
      \vertex [dot, below right=1.4cm of f] (h) {};
      \vertex [right=1.2cm of g] (i) {\(\pi^+\)};
      \vertex [right=1.2cm of h] (j) {\(\pi^-\)};
      \diagram* {
        (a) -- [double] (b),
        (b) -- [fermion] (c),
        (c) -- [fermion] (d),
        (d) -- [fermion] (b),  
        (c) -- [double] (e),         
        (d) -- [double, edge label'=\(f_0\)] (f),
        (f) -- [fermion] (g),
        (h) -- [fermion] (f),
        (g) -- [fermion] (h),
        (g) -- [double] (i),
        (h) -- [double] (j),
      };
     \end{feynman}
    \end{tikzpicture}
  \end{subfigure}%
 \caption{Quark diagrams of the decay $f_1 \to\pi^+ \pi^- \eta$.}
 \label{diagram2}
\end{figure*}
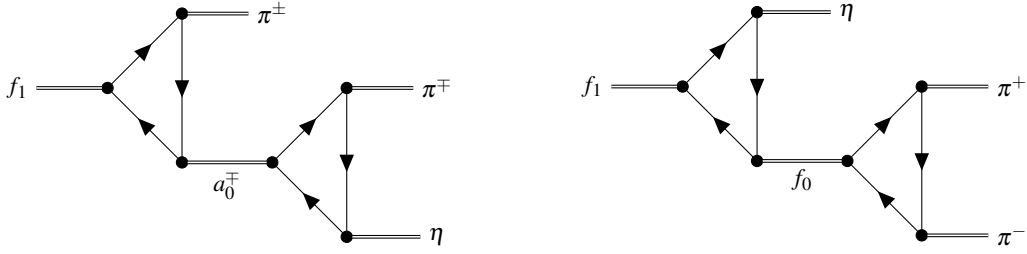%

The decay $f_1 \to \pi\pi\eta$ is described by the contributions of two channels with the intermediate mesons $a_0\pi$ and $f_0 \eta$. The quark diagrams describing the decay $f_1 \to \pi\pi\eta$ are shown in Fig.~\ref{diagram2}. The total amplitude of the considered decay can be represented in the following form:
\begin{eqnarray}
\label{f1pipieta}
\mathcal{M}(f_1 \to \pi^-\pi^+\eta) = i{\left( \mathcal{M}_{a^+_0} + \mathcal{M}_{a^-_0} + \mathcal{M}_{f_0} \right)}^\mu e_\mu^*(p_{f_1}),
\end{eqnarray}
where the contributions of the channels with the intermediate mesons $a^+_0$, $a^-_0$ and $f_0$ are presented in the brackets. The expressions for these channels take the form
\begin{eqnarray}
\mathcal{M}_{a^+_0} = g_{f_1}^u \sqrt{Z_\pi} \cos{\phi} BW_{a^+_0} g_{a^+_0\pi\eta} (p^2_{a^+_0}) {\left( p_{a^+_0} - A_\pi \, p_{\pi^-} \right)}_\mu,
\end{eqnarray}
\begin{eqnarray}
\mathcal{M}_{a^-_0} = g_{f_1}^u \sqrt{Z_\pi} \cos{\phi} BW_{a^-_0} g_{a^-_0\pi\eta} (p^2_{a^-_0}) {\left( p_{a^-_0} - A_\pi \, p_{\pi^+} \right)}_\mu,
\end{eqnarray}
\begin{eqnarray}
\mathcal{M}_{f_0} = \left( g^u_{f_1} \sqrt{Z_\pi} \cos{\phi} \cos{\bar{\theta_\sigma}} \sin{\bar{\theta_\eta}} - \sqrt{2} g^s_{f_1}\sin{\phi} \sqrt{Z_s} \sin{\bar{\theta_\sigma}} \cos{\bar{\theta_\eta}} \right) BW_{f_0} \, g_{f_0\pi\pi} (p^2_{f_0}) {\left( p_{f_0} - A_\eta \, p_{\eta} \right)}_\mu,
\end{eqnarray}
where the factors $A_\pi$ and $A_\eta$ are shown in (\ref{a0pi}) and (\ref{f0eta}), the momenta of the intermediate mesons are defined as $p_{a^+_0} = p_{\pi^+} + p_\eta$, $p_{a^-_0} = p_{\pi^-} + p_\eta$, $p_{f_0} = p_{\pi^+} + p_{\pi^-}$. The functions $g_{a^\pm_0\pi\eta}$ and $g_{f_0\pi\pi}$ describe the decays $a_0 \to \pi \eta$ and $f_0 \to \pi \pi$  
\begin{eqnarray}
g_{a^\pm_0\pi\eta} & = & 4 m_u \sqrt{Z_\pi} g_\eta^u \sin{\bar{\theta_\eta}} \left[ 1- 
\frac{3}{2} \frac{p^2_{a^\pm_0} - M^2_\eta}{M^2_{a_1}} - \frac{3}{2} {\cos{\phi}}^2 \left(1 - \frac{m_s}{m_u}\frac{g_{f_1}^u g_{\eta}^s}{g_{f_1}^s g_{\eta}^u}tg{\phi}ctg{\bar{\theta_\eta}}\right)\frac{p^2_{a^\pm_0} - M^2_\pi}{M^2_{f_1(1285)}} \right.\nonumber\\
&& \left.-\frac{3}{2} {\sin{\phi}}^2 \left(1 + \frac{m_s}{m_u}\frac{g_{f_1}^u g_{\eta}^s}{g_{f_1}^s g_{\eta}^u}ctg{\phi}ctg{\bar{\theta_\eta}}\right)\frac{p^2_{a^\pm_0} - M^2_\pi}{M^2_{f_1(1420)}}\right],
\end{eqnarray}
\begin{eqnarray}
g_{f_0\pi\pi} = 4 m_u \sqrt{Z_\pi} g_\pi \cos{\bar{\theta_\sigma}} \left( 1- 
3\frac{p^2_{f_0} - M^2_\pi}{M^2_{a_1}}
\right).
\end{eqnarray}

The intermediate mesons are described with the Breit-Wigner propagator
\begin{eqnarray}
	BW_{M} = \frac{1}{M_{M}^2 - p^2 - i\sqrt{p^2}\Gamma_{M}},
\end{eqnarray}
where the widths and masses of the intermediate mesons $M_{a_0} = 980$~MeV, $M_{f_0} = 600$~MeV, $\Gamma_{a_0} = 75$~MeV, and $\Gamma_{f_0} = 450$~MeV \cite{ParticleDataGroup:2024cfk}.

In the calculated amplitude, besides the $u$ and $d$ quark loops, the $s$ quark loops have been taken into account when considering the transitions between the axial vector $f_1$ meson and pseudoscalar $\eta$ meson.

As a result, the following estimations of the branching fraction of the decay $f_1 \to \eta \pi\pi$ have been obtained by using the amplitude (\ref{f1pipieta}) in the NJL model
\begin{eqnarray}
	Br(f_1 \to \pi^+\pi^-\eta)_{NJL} = (27.5 \pm 4.13)\%.
\end{eqnarray}

In the case of taking into account only up and down quark loops in the $f_1-\eta$ transitions, the following value of the branching fraction is obtained:
\begin{eqnarray}
	Br(f_1 \to \pi^+\pi^-\eta)_{NJL} = (36.5 \pm 5.48)\%.
\end{eqnarray}

The obtained theoretical estimations are in agreement with the experimental data~\cite{ParticleDataGroup:2024cfk}
\begin{eqnarray}
	Br(f_1 \to \pi^+\pi^-\eta)_{PDG} = (35 \pm 15) \%.
\end{eqnarray}

\section{Decays $f_1 \to \rho \pi$ and $f_1 \to \pi^- \pi^+ \pi^0$}
\label{section::4}
The processes $f_1 \to \rho \pi$ and $f_1 \to \pi^- \pi^+ \pi^0$ occur due to the difference between the $u$ and $d$ quark masses.

\begin{figure*}[t]
 \centering
  \begin{subfigure}{0.5\textwidth}
   \centering
   \begin{tikzpicture}
    \begin{feynman} 
      \vertex (a) {\(f_1 \)};
      \vertex [dot, right=1.2cm of a] (b) {};
      \vertex [dot, above right=1.4cm of b] (c) {};
      \vertex [dot, below right=1.4cm of b] (d) {};
      \vertex [right=1.2cm of c] (g) {\(\pi\)};
      \vertex [right=1.2cm of d] (f) {\(\rho\)};
      \diagram* {
        (a) -- [double] (b),
        (b) -- [fermion] (c),
        (c) -- [fermion] (d),
        (d) -- [fermion] (b),  
        (c) -- [double] (g),         
        (d) -- [double] (f),
      };
     \end{feynman}
    \end{tikzpicture}
  \end{subfigure}%
 \caption{The diagram of the decay $f_1 \to \rho \pi$.}
 \label{diagram_rhopi}
\end{figure*}
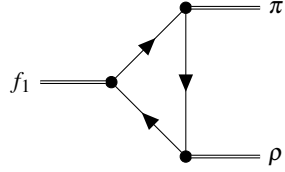%

The diagram of the process $f_1 \to \rho \pi$ is shown in Fig.~\ref{diagram_rhopi}.

The amplitude of the process in the NJL model has the following form:
\begin{eqnarray}
\mathcal{M}(f_1 \to \rho \pi) & = & -i 6 \left(m_d - m_u\right) g_\pi \cos{\phi} e_\mu^*(p_{f_1}) e^\mu(p_{\rho}),
\end{eqnarray}
where $e^\mu(p_{\rho})$ is the polarization vector of the $\rho$ meson.

The branching fraction obtained by using this amplitude is
\begin{eqnarray}
Br(f_1 \to \rho \pi) = (1.9 \pm 0.3) \times 10^{-3}.
\end{eqnarray}

This theoretical estimation for the branching fraction does not contradict the experimental restrictions~\cite{ParticleDataGroup:2024cfk}:
\begin{eqnarray}
Br(f_1 \to \rho \pi)_{PDG} < 3.1 \times 10^{-3}.
\end{eqnarray}

\begin{figure*}[t]
 \centering
  \begin{subfigure}{0.5\textwidth}
   \centering
   \begin{tikzpicture}
    \begin{feynman} 
      \vertex (a) {\(f_1 \)};
      \vertex [dot, right=1.2cm of a] (b) {};
      \vertex [dot, above right=1.4cm of b] (c) {};
      \vertex [dot, below right=1.4cm of b] (d) {};
      \vertex [right=1.2cm of c] (e) {\(\pi^\pm\)};
      \vertex [dot, right=1.2cm of d] (f) {};
      \vertex [dot, above right=1.4cm of f] (g) {};
      \vertex [dot, below right=1.4cm of f] (h) {};
      \vertex [right=1.2cm of g] (i) {\(\pi^\mp\)};
      \vertex [right=1.2cm of h] (j) {\(\pi^0\)};
      \diagram* {
        (a) -- [double] (b),
        (b) -- [fermion] (c),
        (c) -- [fermion] (d),
        (d) -- [fermion] (b),  
        (c) -- [double] (e),         
        (d) -- [double, edge label'=\(\rho^\mp\)] (f),
        (f) -- [fermion] (g),
        (h) -- [fermion] (f),
        (g) -- [fermion] (h),
        (g) -- [double] (i),
        (h) -- [double] (j),
      };
     \end{feynman}
    \end{tikzpicture}
  \end{subfigure}%
 \caption{The diagram of the decay $f_1 \to \pi^- \pi^+ \pi^0$.}
 \label{diagram_f1pipipi}
\end{figure*}
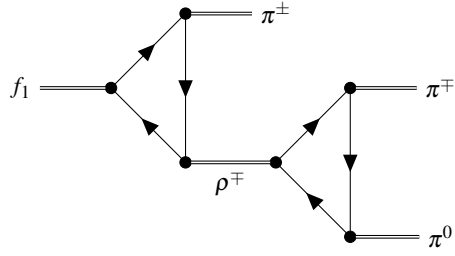

The main channel determining the width of the process $f_1 \to \pi^- \pi^+ \pi^0$ is the vector channel. The diagram of this process is shown in Fig.~\ref{diagram_f1pipipi}.

The appropriate amplitude takes the following form:
\begin{eqnarray}
\mathcal{M}(f_1 \to \pi^- \pi^+ \pi^0) & = & -i 6 \left(m_d - m_u\right) g_\rho g_\pi \cos{\phi}\nonumber\\
&& \times e_\mu^*(p_{f_1}) \left[\left(p_{\pi^0} - p_{\pi^-}\right)^\mu BW_{\rho^-}^{p_{\pi^0} + p_{\pi^-}} + \left(p_{\pi^0} - p_{\pi^+}\right)^\mu BW_{\rho^+}^{p_{\pi^0} + p_{\pi^+}}\right],
\end{eqnarray}
where $p_{\pi^0}$, $p_{\pi^+}$ and $p_{\pi^-}$ are the meson momenta in the final states.

The branching fraction obtained by using this amplitude is
\begin{eqnarray}
Br(f_1 \to \pi^- \pi^+ \pi^0) = (3.2 \pm 0.5) \times 10^{-3},
\end{eqnarray}
and is in agreement with the experimental data~\cite{ParticleDataGroup:2024cfk}
\begin{eqnarray}
Br(f_1 \to \pi^- \pi^+ \pi^0)_{PDG} = (3.0 \pm 0.9) \times 10^{-3}.
\end{eqnarray}

\section{Decays $f_1 \to KK\pi$}
\label{section5}

The decay $f_1 \to K^+K^0\pi^-$ contains the contributions of the channels with the intermediate vector and scalar mesons $K^{*0}$, $K^{*-}$, $a_0$, $K_0^{*0}$ and $K_0^{*-}$. The total amplitude of the decay takes the form
\begin{eqnarray}
\label{kkpi}
\mathcal{M}(f_1 \to K^+K^0\pi^-) = -i 2 \sqrt{2} g_\pi Z_K \, {\left( \mathcal{M}_{K^{*0}} + \mathcal{M}_{K^{*-}} + \mathcal{M}_{a_0} + \mathcal{M}_{K_0^{*0}} + \mathcal{M}_{K_0^{*-}} \right)}_\mu e_\mu(p_{f_1}),
\end{eqnarray}
where the contributions from the channels have the form 
\begin{eqnarray}
\mathcal{M}_{K^{*0}} = \frac{3}{8} \left( (3m_u - m_s) g_{f_1}^u \cos{\phi} + \sqrt{2}(3m_s - m_u) g_{f_1}^s \sin{\phi} \right) BW_{K^{*0}} {\left( a_\pi p_{\pi^-} - a_K \, p_{K^+} \right)}_\mu,
\end{eqnarray}
\begin{eqnarray}
\mathcal{M}_{K^{*-}} = \frac{3}{8} \left( (3m_u - m_s) g_{f_1}^u \cos{\phi} + \sqrt{2}(3m_s - m_u) g_{f_1}^s \sin{\phi} \right) BW_{K^{*-}} {\left( a_\pi p_{\pi^-} - a_K \, p_{K^0} \right)}_\mu,
\end{eqnarray}
\begin{eqnarray}
\mathcal{M}_{a_0} = -(2m_u -m_s) g_{f_1}^u \cos{\phi} \, BW_{a_0} {\left(p_{a_0} - A_\pi \, p_{\pi^-} \right)}_\mu,
\end{eqnarray}
\begin{eqnarray}
\mathcal{M}_{K_0^{*0}} = \frac{1}{2} m_s \left( g_{f_1}^u \cos{\phi} + \sqrt{2} g_{f_1}^s \sin{\phi} \right) BW_{K_0^{*0}} {\left(p_{K_0^{*}} - p_{K^0} \right)}_\mu,
\end{eqnarray}
\begin{eqnarray}
\mathcal{M}_{K_0^{*-}} = \frac{1}{2} m_s \left( g_{f_1}^u \cos{\phi} + \sqrt{2} g_{f_1}^s \sin{\phi} \right) BW_{K_0^{*-}} {\left(p_{K_0^{*}} - p_{K^+} \right)}_\mu,
\end{eqnarray}
where the constants $a_\pi$ and $a_K$ describe the transitions in the decay $K^{*0} \to K^+ \pi^-$
\begin{eqnarray}
a_\pi = 1- \frac{3m_u (3m_u -m_s)}{M^2_{a_1}}, \, a_K = 1- \frac{3m_s (m_u +m_s)}{M^2_{K_{1A}}}.
\end{eqnarray}

By using this amplitude, one can obtain the estimation of the branching fraction of the decay
$Br(f_1 \to K^+K^0\pi^-) = (3.3 \pm 0.5) \%$.
The total width of the decay $f_1 \to KK\pi$ includes the combinations of the mesons in the final state $f_1 \to K^+K^0\pi^-$, $f_1 \to K^-K^0\pi^+$, $f_1 \to K^+K^-\pi^0$ and $f_1 \to K^0\bar{K}^0\pi^0$. As a result, in the framework of the NJL model, for the decay $f_1 \to KK\pi$ one can obtain the estimation $Br(f_1 \to KK\pi)_{NJL} = (9.9 \pm 1.5) \%$. The experimental value of the branching fraction of this process~\cite{ParticleDataGroup:2024cfk}:
\begin{eqnarray}
Br(f_1 \to KK\pi)_{PDG} = (9.0 \pm 0.4) \%.
\end{eqnarray}

\section{Conclusion}
In this paper, we have considered the main strong decays of the axial-vector meson $f_1(1285)$. All calculations were performed in the one-loop quark approximation corresponding to the leading order of expansions in $1/{N_c}$. For the branching fraction of the decay width with the production of the scalar meson $f_1 \to a_0 \pi$, the value $(39.4 \pm 5.91) \%$ was obtained, which is in agreement with the experimental data. In this case, the scalar meson $a_0$ was considered as a chiral-symmetric quark-antiquark partner of the pseudoscalar $\pi$-meson. The decay $f_1 \to \eta \pi^+\pi^-$ is calculated taking into account the intermediate channels with the scalar mesons $a^\pm_0$ and $f_0$. The contribution of the channels with the mesons $a^\pm_0$ to the branching fraction is $15.6 \%$. The contribution of the channel with the scalar meson $f_0$ turns out to be comparable with the channels $Br(f_1 \to a^\pm_0 \pi^\mp\to \eta \pi^+\pi^-) = 7.0 \%$. The total branching fraction partial width taking into account the positive interference of channels $a_0$ and $f_0$ is $Br(f_1 \to \eta \pi^+\pi^-)=27.5 \%$.

From a theoretical point of view, the decay $f_1 \to a_0 \pi$ has recently been considered in \cite{Aceti:2015zva} within the mechanism  with triangle singularity, and the estimated value $Br(f_1 \to \eta \pi^+\pi^-) = 31 \%$, which is in agreement with the experiment and the results of the present work. It should be noted that in our recent work \cite{Volkov:2024wsu} the $\tau \to 3\pi \nu_\tau$ decay was considered taking into account the intermediate channel with the scalar meson $f_0$. In this work, good results were obtained for the spectral function and decay width in agreement with the CLEO data \cite{CLEO:1999rzk}. At the same time, an important role of the $f_0 \pi$ channel contribution to the $\tau \to 3\pi \nu_\tau$ decay width was shown, which improved the agreement with the experiment. 

The suppressed decays $f_1 \to \rho^\pm \pi^\mp$ and $f_1 \to \pi^+\pi^-\pi^0$ were also described. It is shown that these decays occur due to the difference in the masses of the light $u$ and $d$ quarks. In the decay $f_1 \to \pi^+\pi^-\pi^0$ the channel with the charged $\rho^\pm$ mesons completely determines the decay width.

One can obtain the estimation for the branching fraction ratio $Br(f_1 \to \pi^+\pi^-\pi^0)/Br(f_1 \to \pi^+\pi^-\eta) = 1.16 \% \, (0.87\%)$. This estimation within the NJL model agrees with the data of the VES and BES III experiments
\begin{displaymath}
{\frac{Br(f_1 \to\pi^+\pi^-\pi^0)}{Br(f_1 \to\pi^+\pi^-\eta)}} = (0.86 \pm 0.16 \pm 0.20) \%  \cite{Dorofeev:2011zz}, \quad 
{\frac{Br(f_1 \to \pi^+\pi^-\pi^0)}{Br(f_1 \to\pi^+\pi^-\eta)}} = (1.23 \pm 0.55) \%  \cite{BESIII:2015you}.
\end{displaymath}

The obtained results for the decay involving the strange mesons $f_1 \to KK\pi$ are presented in Chapter \ref{section5}. Channels with vector and scalar mesons $K^{*}$, $a_0$ and $K_0^*$ are determined for the decay. An analysis of the decay width $f_1 \to K^+K^0\pi^-$ shows that the contribution of the vector channel is $0.23 \%$. The scalar contributions give $1.9\%$ and the total contribution taking into account the interference is $Br(f_1 \to K^+K^0\pi^-) = 3.3\%$. As a result, for the decay $f_1 \to KK\pi$ taking into account the combination of meson charges in the final state, the resulting partial width $Br(f_1 \to KK\pi) = 9.9\%$ is obtained. This result is in qualitative agreement with the PDG data \cite{ParticleDataGroup:2024cfk}.

The calculations carried out in the present work show that along with various representations of the structure of scalar mesons (see review papers, for example,~\cite{Klempt:2007cp,Pelaez:2015qba,Mai:2022eur}) in the case under consideration the $\bar{q}q$ representation works quite satisfactorily at low energies where the scalar meson is considered as a chirally symmetric partner of the pseudoscalar meson. In the framework of this approach, strong decays of strange scalar mesons $K^*_0(700)$ and $K^*_0(1430)$ \cite{Volkov:2022ukj} have also been recently investigated. In addition, the processes of strange scalar meson production in tau lepton decays \cite{Volkov:2022nxp} were considered.

\subsection*{Acknowledgements}
The authors thank Prof. A. B. Arbuzov for useful discussions. This research has been funded by the Science Committee of the Ministry of Science and Higher Education of the Republic of Kazakhstan Grant No. BR24992891.

\end{document}